\documentclass[prb, preprint, groupedaddress]{revtex4}

\usepackage{amsfonts, amsbsy, amssymb, amsmath, graphicx,  float} 

\usepackage{color}
\usepackage{float}
\restylefloat{figure}
\usepackage{pifont}      
\reversemarginpar

\begin{document}


\title{Phase space barriers and dividing surfaces in the absence of  critical points of the potential energy}



\author{Gregory S. Ezra}
\email[]{gse1@cornell.edu}
\affiliation{Department of Chemistry and Chemical Biology\\
Baker Laboratory\\
Cornell University\\
Ithaca, NY 14853\\USA}

\author{Stephen Wiggins}
\email[]{stephen.wiggins@mac.com}
\affiliation{School of Mathematics\\University of Bristol\\Bristol BS8 1TW\\United Kingdom}


\date{\today}

\begin{abstract}

We consider  the existence of  invariant manifolds in phase space governing reaction dynamics 
in situations where there are no saddle points on the potential energy surface in the relevant 
regions of configuration space. We point out that such situations occur in a number of important 
classes of chemical reactions, and we  illustrate this concretely by considering a  
model for transition state switching in an ion-molecule association reaction 
due to Chesnavich (J. Chem. Phys. {\bf 84}, 2615 (1986)). 
For this model we show that, in the region of configuration space relevant to the 
reaction, there are no saddle points on the  potential energy surface, 
but that in phase space there is a normally hyperbolic invariant manifold 
(NHIM) bounding a dividing surface having the property that the reactive flux through this 
dividing surface is a minimum.  We then describe two methods for finding NHIMs 
and their associated phase space structures in systems with  more than two degrees-of-freedom.
These methods do not rely on the existence of  saddle points, or any other particular feature, 
of the potential energy surface. 

\end{abstract}


\maketitle


\section{Introduction}
\label{sec:intro}

Critical points of the potential energy surface have played, and continue to play, 
a significant role in how one thinks about transformations of physical systems 
\cite{Mezey87,Wales03}. The term `transformation' may refer to chemical 
reactions such as isomerizations 
\cite{Chandler78,Berne82,Davis86,Gray87,Minyaev91,Minyaev94a,Baer96,Leitner99,Waalkens04,Carpenter05,Joyeux05,Ezra09a}
or the analogue of phase transitions for finite size systems \cite{Wales03,Pettini07,Kastner08}.
A comprehensive description 
of this so-called `energy landscape paradigm' is given in ref.\ \onlinecite{Wales03}. 
The energy landscape approach is an attempt to understand  dynamics in the context of the geometrical 
features of the  potential energy surface, i.e., a configuration space approach. 
However,  the arena for dynamics is phase space \cite{Arnold78,AKN88,Wiggins96}, and numerous  
studies of nonlinear dynamical systems have 
taught us that  the rich variety of dynamical behavior possible in nonlinear systems {\em cannot} 
be inferred from geometrical properties of the potential energy surface alone.
(An instructive example is the fact that the well-studied and nonintegrable
H\'enon-Heiles potential can be obtained by series expansion of the 
completely integrable Toda system \cite{Lichtenberg92}.)
Nevertheless, the configuration space based landscape paradigm is physically 
very compelling, and there has been a great deal of work over the past ten years describing 
{\em phase space signatures} of index one saddles \cite{saddle_footnote1}
of the potential energy surface that are relevant to reaction dynamics
(see, for example, refs \onlinecite{ujpyw,Komats3,WaalkensSchubertWiggins08}).
More recently, index two \cite{Heidrich86,Ezra09,haller10} and higher index \cite{Shida05} saddles
have been studied.

The work on index one saddles has shown that, in {\em phase space}, the role of the saddle {\em point} 
is played by an {\em invariant manifold} of saddle stability type, 
a so-called normally hyperbolic invariant manifold or NHIM \cite{Wiggins90,Wiggins94}. 
The NHIM proves to be the anchor 
for the  construction of dividing surfaces that have the  properties of  no (local) recrossing 
of trajectories and minimal (directional) flux \cite{WaalkensWiggins04}.
There is an even richer variety of phase space structures and invariant manifolds associated 
with index two saddles of the potential energy surface, and their implications for 
reaction dynamics are currently under investigation \cite{Ezra09}. 
Fundamental theorems assure the existence of
these phase space structures and invariant manifolds for a range of energy above 
that of the  saddle \cite{Wiggins94}.  
However, the precise extent of this range, as well as the nature 
and consequences 
of any bifurcations of the phase space structures and invariant manifolds 
that might occur as energy is increased, is not known and is a topic of current investigation
\cite{Li09}. 

While work  relating phase space structures and invariant manifolds to saddle 
points on the potential energy surface has provided new insights and techniques for 
studying reaction dynamics \cite{ujpyw,Komats3,WaalkensSchubertWiggins08},  
it certainly does not exhaust all of the  rich possibilities of 
dynamical phenomena associated with reactions. 
In fact, recent work has called into question the 
utility of concepts such as the reaction path and/or 
transition state \cite{Sun02,Townsend04,Bowman06,Lopez07,Heazlewood08,Shepler08}.
Of particular interest for the present work is the recognition that there are 
important classes of chemical reaction,
such as ion-molecule reactions and association reactions in barrierless systems, for which
the transition state is not necessarily
directly associated with the presence of a saddle point on the potential energy surface
(or even the amended potential, which includes centrifugal contributions to the energy 
\cite{Wiesenfeld03,Wiesenfeld05}).
The phenomenon of transition state switching in ion-molecule reactions 
\cite{Chesnavich81,Chesnavich82,Chesnavich86}
provides a good example of the dynamical complexity possible in such systems.

The lack of an appropriate critical point on the potential 
energy surface with which to associate a dividing surface separating
reactants from products in such systems does {\em not} however mean that there are no relevant
geometric structures and invariant manifolds in {\em phase space}. 
In this paper we  discuss the existence of NHIMs, 
along with their stable and unstable manifolds and associated dividing surfaces, 
in regions of phase space that do not correspond to saddle points of the potential energy surface. 
After presenting a simple example motivated by Chesnavich's model for
transition state switching in an ion-molecule
association reaction \cite{Chesnavich86}, we describe
a theoretical framework for describing and computing such NHIMs.
Like the methods associated with index one and two saddles, the method we develop for 
realizing the existence of NHIMs is based on normal form theory; however, rather than normal 
form theory for  saddle-type equilibrium points of Hamilton's equations 
(which are the phase space manifestation of index one and two saddles of the potential energy surface), 
we use normal form theory for  certain hyperbolic invariant tori.  
The  hyperbolic invariant tori (and their stable and unstable manifolds) alone are not adequate, 
in terms of their dimension, for constructing NHIMs that have 
codimension one stable and unstable manifolds (in a fixed energy surface). 
However, by analogy with the use of index one  saddles 
to infer the existence of NHIMs (together with their stable and unstable manifolds, 
and  other  dividing surfaces having appropriate dimensions),  these particular hyperbolic
invariant tori can likewise be used to  infer the existence of phase space structures 
that \emph{are} appropriate for describing  reaction dynamics in situations where there is no 
critical point of the potential energy surface in the relevant region of configuration space. 

Section \ref{sec:example} discusses our simplified version of Chesnavich's model for
transition state switching \cite{Chesnavich86}.  For this 2 DoF system, we exhibit a NHIM (in this case, an
unstable periodic orbit) that is the rigorous dynamical manifestation of the
mininimal flux surface of variational transition state theory \cite{Keck67,Hase83a,Truhlar84}.
In Section \ref{sec:method} we describe a (time-dependent) normal form based approach for 
finding such  NHIMs in phase space. In particular, we  present two variations 
of the method. In Section \ref{sec:method1} we consider systems where 
(to leading order) the system can be 
separated into a two degree-of-freedom (DoF) subsystem and a collection of decoupled ``bath modes''. 
We assume that there exists a hyperbolic periodic orbit in  the 2 DoF subsystem, and 
show that this can be used  to  construct a hyperbolic torus for the  full system. 
We then show that this  hyperbolic torus implies the existence of a NHIM, with stable 
and unstable manifolds that are codimension one in the energy surface.  Appropriate dividing 
surfaces can then be constructed using the  NHIM and the normal form Hamiltonian. 
In Section \ref{sec:method2} we describe a method which requires knowledge 
of the  appropriate hyperbolic invariant torus from the start. The advantage 
of the first method is that it is more intuitive and can exploit the 
considerable number of methods for locating hyperbolic periodic orbits in 
2 DoF Hamiltonian systems. Method two is  more general, but at present 
there are few techniques available for locating hyperbolic invariant tori of 
the appropriate dimension in general $N$ DoF Hamiltonian systems that are {\em not} 
perturbations of integrable systems. 
Section \ref{sec:s&o} concludes.

\newpage 
\section{A motivating example: variational transition state 
for a model barrierless reaction}
\label{sec:example}

\subsection{Introduction}

The conventional approach to variational transition state theory (VTST) for 
barrierless reaction proceeds by
minimizing the reactive flux with respect to variation of some reaction coordinate
chosen \emph{a priori} \cite{Keck67,Hase83a,Truhlar84}.  
The value of the reaction coordinate so determined is therefore 
the location of a flux bottleneck, which is identified with the transition state 
for the particular association reaction.
An invariant phase space characterization of such variationally determined 
dividing surfaces is highly desirable; for $N=2$ DoF, such transition states will 
presumably be associated with unstable 
\emph{periodic orbit dividing surfaces} (PODS) \cite{Pechukas81,Pechukas82,Pollak85}, 
or, more generally,  with NHIMs ($N \geq 2$ DoF)
\cite{Wiggins90,Wiggins94}.

\subsection{Model Hamiltonian}

We consider a highly simplified model for a barrierless association reaction
(cf.\ ref.\ \onlinecite{Chesnavich86}).
The system has 2 DoF: a radial 
coordinate $r$, identified as the reaction coordinate, and a coordinate 
$s$ describing vibrations transverse to
the reaction coordinate.   The radial potential has the character of a 
long-range attractive ion-neutral interaction,
while the potential transverse to the reaction coordinate is harmonic.
The system is nonseparable by virtue of a dependence of the harmonic oscillation 
frequency on the coordinate $r$:
\begin{equation}
\label{ham_1}
H = \frac{p_r^2}{2} + \frac{p_s^2}{2} -\frac{\alpha}{r^4} + \frac{1}{2} \omega(r)^2 s^2.
\end{equation}
We take $\omega(r)$ to have the form
\begin{equation}
\omega(r) = \omega_0  \, e^{-\beta r}
\end{equation}
so that, for $\beta > 0$, the transverse vibration stiffens as $r$ decreases.

A contour plot of the potential
\begin{equation}
\label{pot_1}
v(r, s) = -\frac{\alpha}{r^4} + \frac{1}{2} \omega(r)^2 s^2
\end{equation}
for parameter values $\alpha = 1$, $\beta = 1$, $\omega_0 = 8$ is shown in Fig.\ \ref{plot_1}.

\subsection{Locating the bottleneck}

For the 2 DoF Hamiltonian \eqref{ham_1}, we can compute the \emph{action} of
the transverse vibrational mode as a function of the coordinate $r$ at fixed energy $E$:
\begin{equation}
\label{eq:action_1}
I(r; E) = \left[ E + \frac{\alpha}{r^4}\right]\frac{1}{\omega(r)} 
= \left[ E + \frac{\alpha}{r^4}\right]\frac{e^{\beta r}}{\omega_0}.
\end{equation}

As $r$ decreases, there are two competing tendencies:
\begin{itemize}
\item  Decreasing $r$ increases the amount of energy in the oscillator degree of freedom, 
thereby tending to increase the
action.

\item Decreasing $r$ increases the frequency $\omega(r)$, tending to decrease the action of the
tranverse mode.
\end{itemize}

The competition between these two trends can therefore result in the existence of a \emph{minimum} in
the action as a function of $r$:  see Figure \ref{plot_2}.

The minimum of the action as a function of $r$ corresponds to an extremum of the sum of states
(phase space area) or \emph{flux} as a function of $r$, and hence is interpreted as 
a bottleneck.  In the variational transition state approach, the transition state for 
association is then located at
the value $r=r^\ast$ corresponding to minimum flux.  This bottleneck corresponds to an 
inner or ``tight'' transition state \cite{Chesnavich82,Chesnavich86}.

\subsection{Intrinsic characterization of variational TS: PODS}

The formulation of VTST outlined above for the model association reaction  is
unsatisfactory in that the minimum  flux bottleneck so determined has no intrinsic
dynamical significance. 
It is natural to seek a dynamical, phase space based characterization of the variational TS.
For 2 DoF systems, transition states are identified as PODS
\cite{Pechukas81,Pechukas82,Pollak85}.
The invariant object defining the TS is a  hyperbolic (unstable) periodic orbit; the 1D periodic orbit
forms the boundary of a 2D dividing surface on the 3D energy shell in phase space, 
which is the transition state.
The minimal flux (local no-recrossing) property  of the TS  follows 
from the principle of stationary action \cite{Lanczos86}.

A search using the model potential \eqref{pot_1} reveals the presence of a PODS in the vicinity of
$r \simeq r^\ast$ (the periodic orbit at $E=1$ is shown in Fig.\ \ref{plot_1})
\cite{pods_footnote_1}. 
This PODS is the rigorous dynamical
realization of the variational TS in this simple case.
  
\newpage
\section{Locating NHIMs when there are no (relevant) saddles 
on the potential energy surface}
\label{sec:method}

In this section we describe two methods for locating NHIMs of the type
discussed in the previous section. These methods are inherently phase space approaches,  
based on the existence of a hyperbolic invariant torus 
solution of Hamilton's equations. Normal form theory for hyperbolic invariant tori can be used 
to provide ``good coordinates'' for  computing explicit formulae for   a NHIM,
its stable and unstable manifolds, 
and dividing surfaces in the phase space vicinity of the 
hyperbolic invariant torus on which we base our method, in much the same way that it is 
used to compute similar objects associated with index one saddles of the potential 
energy surface \cite{wwju,ujpyw,WaalkensSchubertWiggins08}.  
A large literature for normal form theory associated with invariant tori 
of Hamilton's equations has been developed over the past  twenty years, 
see ref.\ \onlinecite{bhs96} for an overview, and ref.\ \onlinecite{hanssmann04} for a survey of  the issues  
associated with bifurcation of tori in Hamiltonian systems.
 For our present purposes we use the results contained in ref.\ \onlinecite{Bolotin}, 
 which explicitly discusses the relevant normal form and also clarifies the 
 issue of ``hyperbolicity'' of tori in Hamiltonian systems (concerning which there had previously 
 been some confusion in the literature).

\subsection{Method 1: a relevant  2 DoF subsystem can be identified at leading order}
\label{sec:method1}

Consider  a Hamiltonian of the following form:
\begin{equation}
H = \frac{p_r^2}{2} +  \frac{p_s^2}{2}  + V(r, s) + 
\frac{1}{2} \sum_{i=1}^{n-2} \omega_i \left( u_i^2 + v_i^2 \right) + 
f(r, s, u_1, \ldots , u_{n-2}, p_r, p_s, v_1, \ldots , v_{n-2})
\label{ham1_1}
\end{equation} 	
where $ f(r, s, u_1, \ldots , u_{n-2}, p_r, p_s, v_1, \ldots , v_{n-2})$ 
is at least order $3$, denoted ${\cal O}_3(r, s, u_1, \ldots , u_{n-2}, p_r, p_s, v_1, \ldots , v_{n-2})$.  
In general this  term serves to couple all of the variables, but we 
have written the Hamiltonian in such a way that we can identify a 
clearly defined 2 DoF subsystem, on which we make the following assumption:

\medskip
\noindent
{\bf Assumption:} The  2 DoF subsystem defined by the Hamiltonian:
\begin{equation}
{\cal H} = \frac{p_r^2}{2} +  \frac{p_s^2}{2}  + V(r, s),
\label{hamsub1_1}
\end{equation}
has a  hyperbolic periodic orbit, denoted ${\cal P} = (r(t), s(t), p_r (t), p_s (t))$.

To construct a NHIM for \eqref{ham1_1} we proceed as follows.

\medskip
\noindent
{\bf Step 1: Transform the 2 DoF subsystem to normal form in a neighborhood of the periodic orbit.}

Following ref.\ \onlinecite{Bolotin},  we can find an invertible transformation $T_0$ 
(as smooth as the Hamiltonian) defined in a neighborhood of the periodic orbit 
\begin{subequations} 
\label{trans_per}
\begin{align}
T_0  & :  (r, s, p_r, p_s) \mapsto T_0 (r, s, p_r, p_s) \equiv (I, \theta, x, y) \\
T_0^{-1} & : (I, \theta, x, y) \mapsto T_0^{-1} (I, \theta, x, y) \equiv  (r, s, p_r, p_s)
\end{align}
\end{subequations}
such that
the  2 DoF Hamiltonian takes the form:
\begin{equation}
K = \omega I + \lambda xy + {\cal O}_2(I) +  {\cal O}_3 (I, x, y), 
\label{nformsub1_1}
\end{equation}
where we  can take $\omega, \, \lambda >0$. Of course, nontrivial calculations 
are required  in going from \eqref{hamsub1_1} to \eqref{nformsub1_1}.
In particular, after the hyperbolic periodic orbit is located, a time-dependent 
translation to  ``center'' the coordinate system on the periodic orbit must be carried out; 
the resulting Hamiltonian is then Taylor expanded  about the origin (i.e., the periodic orbit),  
a Floquet-type transformation constructed to make (to leading order) the  
dynamics  in the normal direction to the periodic orbit constant 
(i.e., $\lambda$ is constant in \eqref{nformsub1_1}), then, finally,  
Hamiltonian normal form theory  is applied to the result. 
Details of the methodology for carrying out this procedure for specific examples are described in 
refs \onlinecite{jv, jorba99}. 
For the purposes of demonstrating the existence of a NHIM, we only need to know that 
such transformations can in principle be carried out. 

\noindent
{\bf Step 2: Use the normal form transformation for the 2 DoF subsystem to rewrite \eqref{ham1_1}.}

We use the normal form transformation of the 2 DoF subsystem to express \eqref{ham1_1} as follows:
\begin{multline}
\bar{H}= \omega I + \lambda xy +  
\frac{1}{2} \sum_{i=1}^{n-2} \omega_i \left( u_i^2 + v_i^2 \right) 
\\ + F( u_1, \ldots , u_{n-2}, v_1, \ldots , v_{n-2}, I, \theta, x, y) 
+ {\cal O}_2(I) + {\cal O}_3 (I, x, y)
\label{ham1_2}
\end{multline}
where
\begin{multline}
F( u_1, \ldots , u_{n-2}, v_1, \ldots , v_{n-2}, I, \theta, x, y)  = \\ 
f(r (I, \theta, x, y), s(I, \theta, x, y), u_1, \ldots , u_{n-2}, 
p_r(I, \theta, x, y), p_s(I, \theta, x, y), v_1, \ldots , v_{n-2}).
\end{multline}

\medskip
\noindent
{\bf Step 3: Use Hamiltonian \eqref{ham1_2} to conclude the existence of a NHIM.}

Let
\begin{equation}
I = \frac{1}{2} (w^2 + z^2), \, \theta = \tan^{-1} \left(\frac{z}{w}\right).
\end{equation}
We then rewrite \eqref{ham1_2} as 
\begin{align}
\label{ham1_3}
& {\cal H}  =   \lambda xy + \frac{\omega}{2} (w^2 + z^2) +  
\frac{1}{2} \sum_{i=1}^{n-2} \omega_i \left( u_i^2 + v_i^2 \right) \nonumber \\
& +  F( u_1, \ldots , u_{n-2}, v_1, \ldots , v_{n-2}, I (w, z), \theta (w, z), x, y) + {\cal O}_2(I(w, z)) + 
{\cal O}_3 (I(w, z), x, y). 
\end{align}
By construction, ${\cal H} (0) =0$.  Neglecting higher order terms in \eqref{ham1_3}, we obtain:
\begin{equation}
{\cal H}_{\rm trunc}   =   \lambda xy + \frac{\omega}{2} (w^2 + z^2) +  
\frac{1}{2} \sum_{i=1}^{n-2} \omega_i \left( u_i^2 + v_i^2 \right).
\label{ham1_4}
\end{equation}

\noindent

If we set $x=y=0$, then
on the energy surface ${\cal H}_{\rm trunc} = h>0$, 

\begin{equation}
\frac{\omega}{2} (w^2 + z^2) +  \frac{1}{2} \sum_{i=1}^{n-2} \omega_i \left( u_i^2 + v_i^2 \right)=h>0,
\label{nhim1_1}
\end{equation}

\noindent
is a normally hyperbolic invariant $2n-3$ sphere in the $2n$-dimensional space with coordinates 
$( u_1, \ldots , u_{n-2}, v_1, \ldots , v_{n-2}, w, z, x, y)$, having $2n-2$ 
dimensional stable and unstable manifolds in the $2n-1$ dimensional energy surface.
The persistence theory for NHIMs implies that this 
manifold persists when the higher order terms are added for energies sufficiently close to $h=0$
(cf.\ ref.\ \onlinecite{Wiggins94}).

At this point we are in a position where  normal form theory can be used  on \eqref{ham1_3}  
to construct a new set of coordinates (the normal form coordinates) in which \eqref{ham1_3} 
assumes a particularly simple form that results in explicit formulae for the NHIM, 
its stable and unstable manifolds, and dividing surfaces between 
regions of the phase space corresponding to reactants and products. 
The normal form algorithm also provides the transformation from the original 
physical coordinates and its inverse, and this allows us to map these surfaces 
back into the original physical coordinates, as described in ref.\ 
\onlinecite{WaalkensSchubertWiggins08}.

\subsection{Method 2: A hyperbolic torus of dimension $n-1$ can be located in an $n$ DoF system.}
\label{sec:method2}

The advantage of method 1 is that it can make use of extensive prior work on 
locating  periodic orbits, and determining their stability, in 2 DoF systems.  
In method 2 the starting point is knowledge of the existence of a hyperbolic  
torus of dimension $n-1$ in the $2n$-dimensional phase space, 
with the frequencies on the torus satisfying a diophantine condition \cite{Schmidt80}.
In this case, it follows from ref.\ \onlinecite{Bolotin} that an invertible 
transformation of coordinates, valid in a neighborhood of the torus, 
can be found where the system has the following form:
\begin{equation}
H = \omega_1 I_1 + \ldots + \omega_{n-1} I_{n-1} + \lambda xy + 
{\cal O}_2(I_1, \ldots , I_{n-1} ) + {\cal O}_3 (I_1, \ldots , I_{n-1}, x, y).
\label{hamnf2_1}
\end{equation}
Now let
\begin{equation}
I_i = \frac{1}{2} (u_i^2 + v_i^2),  \, \theta_i = \tan^{-1} \left(\frac{v_i}{u_i}\right), 
\, i=1, \ldots, n-1.
\end{equation}
In terms of these coordinates the Hamiltonian \eqref{hamnf2_2} has the form:
\begin{equation}
{\bar H} = \frac{1}{2} \sum_{i=1}^{n-1}\omega_i (u_i^2 + v_i^2) +  \lambda xy + {\cal O}_2(u_1^2 + v_1^2, \ldots, u_{n-1}^2 + v_{n-1}^2 ) + {\cal O}_3 (u_1^2 + v_1^2, \ldots, u_{n-1}^2 + v_{n-1}^2, x, y)
\label{hamnf2_2}
\end{equation}

We now proceed exactly as for method 1.  
Neglecting the terms in \eqref{hamnf2_2} of order 3 and higher gives:
\begin{equation}
{\bar H_{\rm trunc}} = \frac{1}{2} \sum_{i=1}^{n-1} \omega_i (u_i^2 + v_i^2) +  \lambda xy. 
\label{hamnf2_3}
\end{equation}
If we set $x=y=0$, then
on the energy surface ${\bar H_{\rm trunc}} =h >0$,
\begin{equation}
\frac{1}{2} \sum_{i=1}^{n-1}\omega_i (u_i^2 + v_i^2) =h>0,
\label{hamnf2_4}
\end{equation}
is a normally hyperbolic invariant $2n-3$ sphere in the $2n$-dimensional space with coordinates 
$( u_1, \ldots , u_{n-1}, v_1, \ldots , v_{n-1}, x, y)$ 
having $2n-2$ dimensional stable and unstable manifolds in 
the $2n-1$ dimensional energy surface. The persistence theory for 
NHIMs implies that this  
manifold persists when the higher order terms are added for energies sufficiently close to $h=0$.

Method 2 relies on first finding an appropriate hyperbolic invariant torus, 
and in general the existence of this object will need to be verified numerically. 
There has been a great deal of activity developing such numerical methods in recent years. 
See, for example, refs 
\onlinecite{dieci91, dieci95, edoh00, gabern05, schilder05, haro06, dankowicz06, jorba09}.

Finally, it is also worth noting the difference between the $h=0$ limit for index $1$ 
saddles  on the potential energy surface compared with the torus case in phase space. 
In the former case the NHIM shrinks down to a point on the potential energy surface 
(i.e. configuration space),
while in the latter case it shrinks down to a torus in phase space.

\newpage
\section{Summary and conclusions}
\label{sec:s&o}

In this paper we have exhibited 
a normally hyperbolic invariant manifold (NHIM \cite{Wiggins94}, in this
case a PODS \cite{Pechukas81,Pechukas82,Pollak85}, or unstable periodic orbit)
defining a flux bottleneck in a simple model of an ion-molecule reaction, 
for which  there is no associated critical point of the 
potential energy surface.  We have also developed a theoretical framework
for showing the existence of such NHIMs. Two methods were
described that are in principle
suitable for computing such phase space objects in the multidimensional case.

For index one saddles a software package has been developed that allows one to compute 
the normal form associated with the corresponding saddle-center-$\ldots$-center stability type 
equilibrium point to high order for  multi-dimensional systems, with control over the accuracy
\cite{software_footnote_1}.  
Accuracy is assessed by a battery of tests, and specifying an accuracy may affect the order of 
the normal form that can be computed as well as the dimensionality of the system that can be treated. 
In general, these issues must be analyzed on a case-by-case basis.  Nevertheless, the normal form, 
and most importantly the transformation from the original physical coordinates and its inverse, 
allow us to  realize the NHIM, its stable and unstable manifolds, and dividing surfaces between 
regions of the phase space corresponding to reactants and products.  Moreover,  
flux through the dividing surfaces can be computed as an integral over the NHIM, and the normal 
form coordinates provide a  natural way of selecting distributions of initial conditions of 
trajectories on the dividing surfaces to compute gap times \cite{thiele, Ezra09a}.  
It should be possible to develop similar 
software for computing normal forms associated with hyperbolic tori of the type discussed 
above, and one expects 
that the  normal form will, similarly, allow one to realize phase space structures relevant to 
reaction dynamics as well as to compute fluxes and sample distributions of initial conditions. 

This program has yet to be carried out, but 
the essential computational elements of the approach can be found in refs \onlinecite{jv, jorba99}. 
Such capabilities would be very useful for the study of reaction dynamics in multimode
systems exhibiting transition state switching, for example ref.\ \onlinecite{Chesnavich82}.

\acknowledgments

SW  acknowledges the support of the  Office of Naval Research Grant No.~N00014-01-1-0769.
GSE and SW both acknowledge the stimulating environment of the NSF sponsored Institute for
Mathematics and its Applications (IMA) at the University of Minnesota,
where this work was begun.


\begin{thebibliography}{10}
\bibitem{pods_footnote_1}
\bibinfo{note}{{The pods at energy $E$ for the reversible Hamiltonian
  \eqref{ham_1} is located by initiating trajectories with zero velocity on the
  equipotential line $V=E$. The initial position is varied until a
  self-retracing (periodic) orbit is obtained. For the 2 DoF system studied,
  this is a 1-parameter search.}}
\end{thebibliography}

\def\cprime{$'$}

\newpage

\section*{Figure captions}

 \begin{figure}[H]
  \caption{\label{plot_1}  Contour plot of the model potential of eq.\ \eqref{pot_1}. 
   Parameter values $\alpha = 1$, $\beta = 1$, $\omega_0 = 8$.  Superimposed on the potential contours
   is a PODS defining a ``tight'' transition state, computed at energy $E=1$.}
  \end{figure}

 \begin{figure}[H]
  \caption{\label{plot_2}  Action $I(E, r)$ (eq.\ \eqref{eq:action_1}) as a function of $r$, 
  computed at energy $E=1$.}
  \end{figure}

 \newpage
 
 \begin{figure}[H]
  \centering
  \includegraphics[width=4.5in]{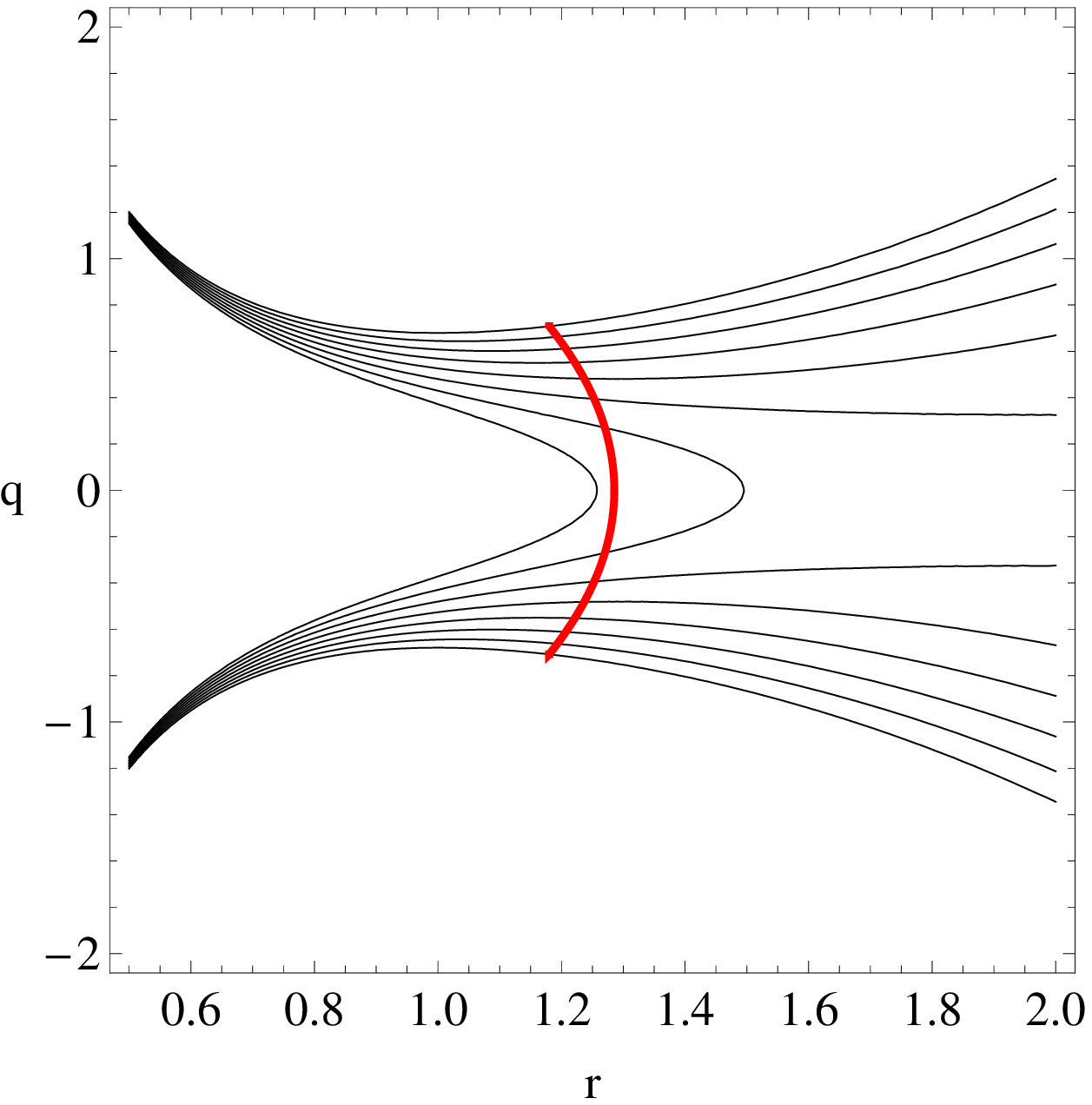}\\
  \vspace*{0.4in}
  \end{figure}

   \vspace*{1.5cm}
   FIGURE 1

 \newpage
 
 \begin{figure}[H]
  \centering
  \includegraphics[width=4.5in]{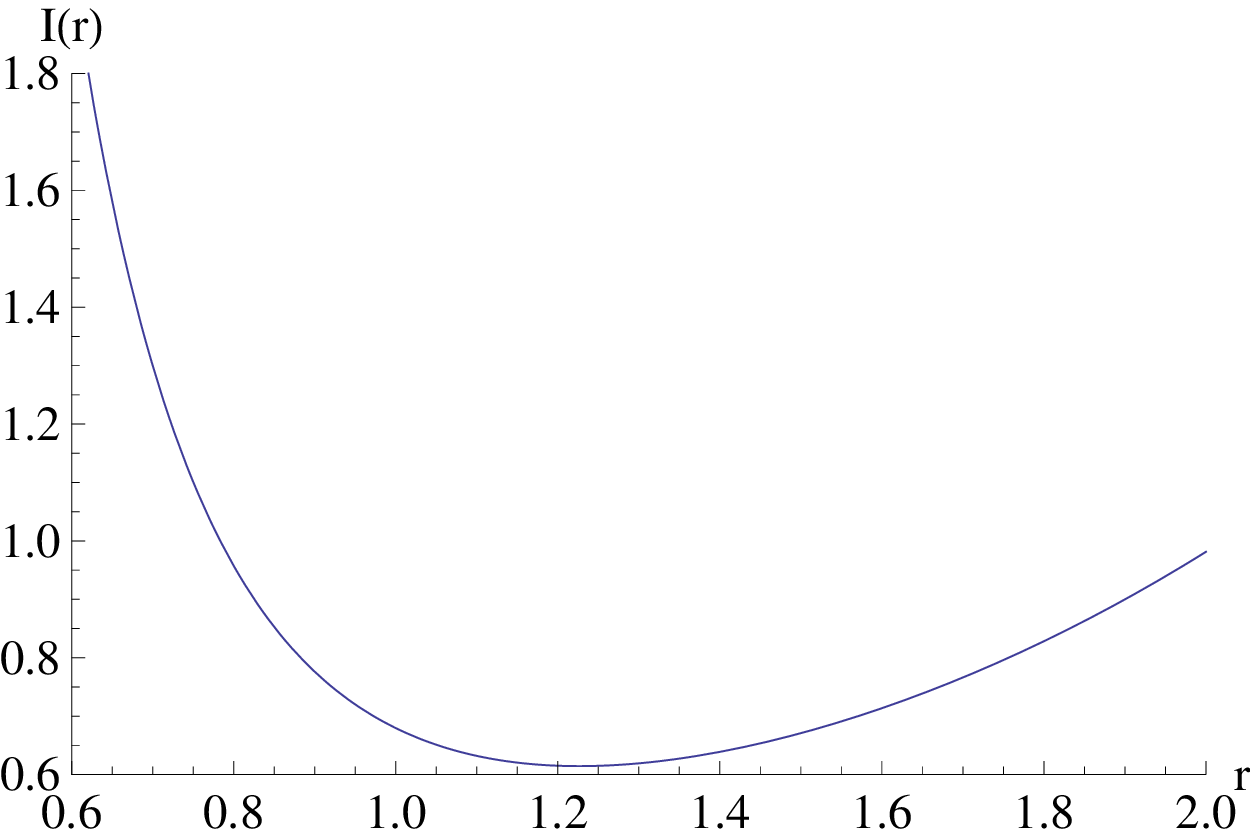}\\
   \vspace*{0.4in}
  \end{figure}

     \vspace*{1.5cm}
     FIGURE 2

\end{document}